\documentstyle[aps]{revtex} 
\begin{document}
\draft
\title{The Tolman VII solution, trapped null orbits and w - modes }
\author{Nicholas Neary, Mustapha Ishak and Kayll Lake\cite{email}}
\address{Department of Physics, Queen's University, Kingston, Ontario, Canada, K7L 3N6 }
\date{\today}
\maketitle
\begin{abstract}
The Tolman VII solution is an exact static spherically symmetric perfect fluid solution of Einstein's equations that exhibits a surprisingly good approximation to a neutron star. We show that this solution exhibits trapped null orbits in a causal region even for a tenuity (total radius to mass ratio) $> 3$. In this region the dynamical part of the potential for axial w - modes dominates over the centrifugal part.
\end{abstract} 
\section{Introduction} Exact solutions of Einstein's equations provide both a route to the physical understanding of relativistic phenomena and a check on numerical procedures. Unfortunately, even in the case of isolated static spherically symmetric perfect fluids, very few solutions are known which pass even elementary tests of physical relevance \cite{del}. In a recent study \cite{icnl} four types of ostensibly reasonable solutions were examined for the internal trapping of null geodesics \cite{trap} along with the associated resonance scattering of axial gravitational waves \cite{CHANDRAFER}, \cite{vishu}, \cite{unphysical}. The solutions examined in \cite{icnl} have a large abrupt density discontinuity (``phase transition") at the boundary \cite{phase}. As a consequence, though physically reasonable, they are not good models of (say) neutron star structure \cite{lattimer}. There are only two known exact static spherically symmetric perfect fluid solutions which match onto vacuum without any density discontinuity at the boundary: the Buchdahl solution  \cite{buch}, and a special case of the Tolman VII solution \cite{tolman}. These solutions are both quite reasonable models of neutron star structure \cite{lattimer}. The Buchdahl solution does not exhibit internal trapping of null geodesics \cite{icnl}. In this paper we show that the Tolman VII solution exhibits trapped null orbits even for a tenuity (total radius to mass ratio) $> 3$. In this region the dynamical part of the potential for axial w - modes actually dominates over the centrifugal part.

\section{The Tolman VII Solution}
\subsection{Outline of the solution}
We begin with the static spherically symmetric metric in ``curvature" coordinates\cite{units}
\begin{equation}
ds^2 = ds^2_{\Gamma} + r^2 d\Omega^2, \label{static}
\end{equation}
where $d\Omega^2$ is the metric of a unit sphere and $d^2_{\Gamma}$ is a static Lorentzian 2-surface with coordinates $(t,r)$. The source of (\ref{static}), by way of Einstein's equations, is considered to be a perfect mathematical fluid generated by radial streamlines of constant $r$. The solution terminates and matches onto a Schwarzschild vacuum upon the vanishing of the isotropic pressure. 

The Tolman VII solution begins with the ansatz
\begin{equation}
\frac{8 \pi \rho(r)}{3} = \frac{1}{\beta^2K^2T}(\beta^2T-(\frac{r}{K})^2) \label{density}
\end{equation}
where $\rho(r)$ is the energy density and $\beta$, T and $K$ are constants. The physical meaning of these constants is clear at the outset: $K^2 = \frac{3}{8 \pi \rho_{o}}$ , where $\rho_{o}$ is the central density, and $\rho = \frac{(T-1)}{K^2T}$ for $ \frac{r}{K} = \beta$. The isotropic pressure ($p$) vanishes at $\frac{r}{K} = \beta$ defining the boundary ($\Sigma$). (The Schwarzschild interior solution is not the $\beta \rightarrow \infty$ limit.) Physically reasonable values of $T$ for neutron stars are $T=1+\epsilon$ where $\epsilon <\sim 10^{-8}$. In this paper, without any change to our conclusions, we take $T=1$. It is convenient to use the dimensionless variable $x \equiv \frac{r}{K}$.
 
With (\ref{density}) the effective gravitational mass ($m$) is given by
\begin{equation}
m(\beta,x) = K\frac{x^3}{10}(5-\frac{3}{\beta^2}x^2) \label{mass}
\end{equation}
which gives a tenuity $\frac{r}{m(r)}|_{\Sigma} = \frac{5}{\beta^2} \equiv \alpha$.
In contrast to (\ref{density}) and (\ref{mass}), the isotropic pressure is not a simple function. (Indeed, this caused Tolman to abandon the solution \cite{tolman}.) The pressure is given in the Appendix. We record the metric:
\begin{equation}
ds^2_{\Gamma} = -D^2sin(ln(\sqrt{\frac{\sqrt{a(\beta,x)}+b(\beta,x)}{C(\beta)}}))^2dt^2+\frac{dr^2}{a(\beta,x)} ,\label{metric}
\end{equation}
where
\begin{equation}
a(\beta,x)=1-x^2+\frac{3x^4}{5\beta^2}, \label{a}
\end{equation}
\begin{equation}
b(\beta,x)= \frac{1}{\beta}\sqrt{\frac{3}{5}}(x^2-\frac{5\beta^2}{6}), \label{b}
\end{equation}
and $D$ defines the choice of scale for $t$ (which, of course, is of no physical consequence). The number $C(\beta)$ (which is governed by $T$) is given in the Appendix. Without regard to causality, the allowed range in $\beta$ is $0<\beta < \sim 1.3896$ where the upper bound is set by the requirement of a finite central pressure.
\subsection{Notation}
We simply review here some key equations to set the notation (see e.g. \cite{icnl} for details). Non-radial null geodesics satisfy 
\begin{equation}
r^{2}r^{\bullet2}=(1-\frac{2m(r)}{r})(\frac{B(r)^{2}}{b^{2}}-1),\label{nrc}
\end{equation}
with
\begin{equation}
B(r) \equiv \frac{r}{\sqrt{g_{tt}}},  \label{nrd}
\end{equation}
where $ \bullet \equiv d/d\lambda$ for affine $\lambda$, and $b$ is a constant $>0$, the ``impact parameter". The evolution of null geodesics is restricted by the condition $b \leq B(r)$. The odd parity (axial) w-modes are non-radial perturbations of the spacetime which do not couple to the fluid at all. In terms of the frequency $\varpi$ and mode number $l \geq 2$ the governing equation is given by \cite{CHANDRAFER}
\begin{equation}
	(\frac{d^2 }{d{r_*}^2}+\varpi^2)Z=V(r_*) Z , \label{wave}
\end{equation}
where $r_*$ is the ``tortoise" coordinate. The potential is conveniently expressed in terms of $r$ and is given by
\begin{equation}
	V(r)=\frac{1}{B(r)^2}(l(l+1)+4\pi r^2(\rho(r)-p(r))-\frac{6 m(r)}{r}). \label{potential}
\end{equation}
 The ``centrifugal" part of the potential is given by $\frac{(l(l+1)}{B(r)^2}$, the remainder being the ``dynamical" part. A necessary condition for the occurrence of resonance scattering of axial gravitational waves by an isolated distribution of fluid is a local minimum in $V(r)$ within the boundary of the fluid.
\subsection{Properties}
We start by noting that for an appropriate choice of the constant $D$ in metric (\ref{metric}) the potential impact parameter $B(r)$ and associated potential $V(r)$ match continuously onto their exterior Schwarzschild vacuum values at the boundary $\Sigma$. The appropriate value of $D$ is shown in Figure 1 where, for convenience, we also record the tenuity $\frac{r}{m(r)}|_{\Sigma} \equiv \alpha$. The choice for $D$ is of no physical consequence. We have included Figure 1 simply to emphasize the fact that the continuous matching of $B(r)$ and $V(r)$ is possible. Figure 2 shows the value of $\frac{r}{\beta K}$ for which $B(r)$ reaches a local maximum compared to the value of $\frac{r}{\beta K}$ below which the classical adiabatic sound speed ($\textrm{v}_s^2 \equiv \frac{dp}{d\rho}$) becomes superluminal. We conclude that $\beta < \sim 1.29663$ and that the tenuity $> \sim 2.97402$ for the trapping of null geodesics in a causal ($\textrm{v}_s<1$) region. Figure 3 records some gross features of the potential impact parameter $B(r)$. The minimum $\beta$ for which there is a local maximum in $B(r)$ is $\beta \sim 1.24955$ corresponding to a maximum tenuity of $ \sim 3.20229$. The Tolman VII solution therefore exhibits trapped null orbits for a tenuity $>3$.  This is the first exact causal solution which has been shown to have this property. As regards the full potential $V(r)$, we find that the minimum $\beta$ for which there is a local minimum in $V(r)$ is $\beta \sim 1.28582$ corresponding to a maximum tenuity of $ \sim 3.02419$. This means that the dynamical part of the potential for axial w - modes in fact dominates over the centrifugal part in the range of tenuity  $ \sim 3.02419 < \alpha < \sim 3.20229$. This is examined in Figure 4. Some gross features of the potential $V(r)$ are shown in Figure 5. For completeness, a comparison of the Tolman VII solution with the well known (and widely used) Schwarzschild interior solution is given in Figure 6.
\section{Discussion}
The Tolman VII solution is the first exact causal solution which has been shown to exhibit trapped null orbits for a tenuity $>3$. Moreover, this solution shows that the dynamical part of the potential for axial w - modes can in fact dominate over the centrifugal part over a  range in tenuity. These features would not be expected from any studies of the Schwarzschild interior solution. Whereas the potential $V(r)$ satisfies the necessary condition for the existence of trapped w-modes even for a tenuity $>3$, it seems unlikely that trapped w-modes actually exist in this solution for tenuities in this range. Nonetheless, this solution makes it clear that trapped null orbits and w-modes exist in causal solutions with realistic equations of state and everywhere smooth densities. There is reason to believe then that the purely relativistic phenomena of trapped null orbits and w-modes are part of real astrophysical systems and not just curiosities of toy models.

\section*{Acknowledgments}
This work was supported by a grant (to KL) from the Natural Sciences and Engineering Research Council of Canada. 
Portions of this work were made possible by use of \textit{GRTensorII}\cite{grt}. We would like to thank Don Page for several very helpful comments on the Tolman VII solution.

\begin{appendix}
\section{Isotropic Pressure}
The pressure can be given in the form
\begin{equation}
8 \pi p(\beta,x) = \frac{1}{x^2K^2}(\frac{2 x^2 \sqrt{a(\beta,x)}}{\beta}\sqrt{\frac{3}{5}}cot(ln\sqrt{\frac{\sqrt{a(\beta,x)}+ b(\beta,x)}{C(\beta)}})+a(\beta,x)-1), \label{pressure}
\end{equation}
where $a(\beta,x)$ and $b(\beta,x)$ are given by (\ref{a}) and (\ref{b}) respectively and $C(\beta)$ is a constant (determined by the condition $T=1$) given by 
\begin{equation}
C(\beta) = \frac{1}{\sqrt{15}}exp(2arctan\Psi(\beta))(\delta+\frac{\beta}{2}), \label{c}
\end{equation}
where
\begin{equation}
\delta \equiv \sqrt{15-6\beta^2}
\end{equation}
and
\begin{equation}
\Psi(\beta) \equiv \frac{23\beta^2 \delta+24 \beta^3-60(\beta + \delta)}{\beta(-23 \beta^2+60+4 \beta \delta)}.
\end{equation}
Note that $p(x=\beta) = 0$.

\end{appendix}

\newpage
\begin{figure}
\caption{The abscissa is $\beta$. The upper curve gives the value of $D^2$ for which the potential impact parameter $B(r)$ and associated potential for axial w - modes $V(r)$ match continuously onto their exterior Schwarzschild vacuum values at the boundary $\Sigma$ ($r=\beta K$). The scale for $D^2$ is on the left. The lower curve gives the tenuity $\frac{r}{m(r)}|_{\Sigma} \equiv \alpha$ for which the scale is on the right.} \label{Dandalpha} 
\label{fig1}
\end{figure}
\begin{figure}
\caption{The ordinate is $\frac{r}{\beta K}$ and the abscissa $\beta$. The curve uppermost on the left gives the value of $\frac{r}{\beta K}$ for which the potential impact parameter $B(r)$ reaches a local maximum. The other curve gives the value of $\frac{r}{\beta K}$ below which the sound speed becomes superluminal. The curves cross at $\beta \sim 1.29663$ at a tenuity $\sim 2.97402$} \label{Bsound} 
\label{fig2}
\end{figure}
\begin{figure}
\caption{Gross features of the potential impact parameter $B(r)$. The vertical scale is irrelevant. The abscissa is $\frac{r}{\beta K}$. The minimum $\beta$ for which there is a local maximum in $B(r)$ is $\beta \sim 1.24955$ corresponding to a maximal tenuity of $ \sim 3.20229$. (a) The values of $\beta$ and tenuity top to bottom are $(1.29663,2.97402),(1.29,3.005),(1.28,3.05176),(1.27,3.1),(1.26,3.14941),(1.25,3.2)$. (b) An enlarged view of the lowest curve in (a).} \label{Bgross} 
\label{fig3}
\end{figure}
\begin{figure}
\caption{Comparison of the potential impact parameter $B(r)$ and the potential $V(r)$. The abscissa is $\frac{r}{\beta K}$. The vertical scale is irrelevant. (a) Dominance of the dynamical part of the potential. Here $\beta = 1.28$. The potential $V(r)$ (upper curve on the left) is monotone decreasing yet the potential impact parameter $B(r)$ reaches a local maximum in a causal region. (b) Here $\beta = 1.29663$ and $B(r)$ (upper curve on the left) has a local maximum where the sound speed becomes luminal. } \label{bandv} 
\label{fig4}
\end{figure}
\begin{figure}
\caption{Gross features of the potential $V(r)$. The abscissa is $\frac{r}{\beta K}$. The vertical scale is irrelevant.  The minimum $\beta$ for which there is a local minimum in $V(r)$ is $\beta \sim 1.28582$ corresponding to a maximal tenuity of $ \sim 3.02419$. (a) The values of $\beta$ and tenuity top to bottom are $(1.289,3.00929), (1.29,3.00463), (1.291,2.99997),(1.292,2.99533), (1.293,2.99070)$, $(1.294,2.98608), (1.295,2.98147), (1.296,2.97687), (1.2966,2.97415)$. (b) An enlarged view of the uppermost curve in (a).} \label{v} 
\label{fig5}
\end{figure}
\begin{figure}
\caption{The Tolman VII solution compared to the Schwarzschild interior solution. In each case the abscissa is a dimensionless measure of distance to the boundary $\Sigma$ (set at $1$).  The vertical scale is irrelevant. The sound velocity is subluminal throughout the range shown for the Tolman VII solution (and infinite for the Schwarzschild interior solution). (a) The potential impact parameter $B(r)$ for the Tolman VII solution (upper curve) compared to the Schwarzschild interior solution. The tenuity for the Tolman VII solution is 3.052 compared to 2.632 for the Schwarzschild solution. (b) The potential $V(r)$ for the Tolman VII solution (lower curve) compared to the Schwarzschild interior solution. The tenuity for the Tolman VII solution is 2.974 compared to 2.632 for the Schwarzschild solution.} \label{schtolmanbv} 
\label{fig5}
\end{figure}


\begin{references}
\bibitem{units} We use geometrical units throughout. For example, the tenuity $r_{\Sigma}/M$ is actually $c^2r_{\Sigma}/GM$.
\end{references}
\end{document}